\documentclass[10pt]{article}

\usepackage{marvosym}

\usepackage{amssymb,amsmath,amsthm}
\usepackage{dsfont}
\usepackage{enumerate}
\usepackage{epsfig}
\usepackage{version}
\usepackage{fullpage}
\usepackage{url}
\usepackage[english]{babel}
\usepackage[utf8]{inputenc}
\usepackage{cmbright}%police plus sobre
\usepackage{eurosym}
\usepackage{hyperref}
\usepackage{amssymb}
\usepackage{subcaption}
\usepackage{color}

\DeclareMathOperator\RR{\mathbb{R}}

\theoremstyle{plain}
\newtheorem{theorem}{Theorem}
\newtheorem{lemma}[theorem]{Lemma}

\newtheorem{proposition}[theorem]{Proposition}
\newtheorem{claim}[theorem]{Claim}

\theoremstyle{definition}

\date{}

\newenvironment{proof-}{\par \noindent \textbf{Proof.} }{\hfill$\Box$\medskip}

\title{Homothetic triangle representations of planar graphs}
\author{Daniel Gonçalves, Benjamin L\'ev\^eque, Alexandre Pinlou}

\begin{document}
\maketitle

\begin{abstract}
We prove that every planar graph is the intersection graph of
homothetic triangles in the plane\footnote{This result was already
  annouced in~\cite{GLP11}}.
\end{abstract}

\section{Introduction}

Here, an \emph{intersection representation} is a collection of
shapes in the plane. The \emph{intersection graph}
described by such a representation has one vertex per
shape, and two vertices are adjacent if and only if the corresponding
shapes intersect.
In the following we only consider shapes that are homeomorphic to
disks.  In this context, if for an
intersection representation the shapes are interior disjoint, we call
such a representation a \emph{contact representation}. In such a
representation, a \emph{contact point} is a point that is in the
intersection of (at least) two shapes.

Research on contact representations of (planar) graphs with predefined
shapes started with the work of Koebe in 1936, and was recently widely
studied; see for example the literature for
disks~\cite{And70,CdV91,Koe36}, triangles~\cite{FOR94}, homothetic
triangles~\cite{GHK10,GLP11,KKLS06,Swg17},
rectangles~\cite{Fel12,T84}, squares~\cite{Lov,Schramm93},
pentagons~\cite{Fel17}, hexagons~\cite{GHKK10}, convex
bodies~\cite{S90}, or (non-convex) axis aligned
polygons~\cite{ABF13,GIP18}. In the present article, we focus on
homothetic triangles. It has been shown that many planar graphs admit
a contact representation with homothetic triangles. 

\begin{figure}[h]
  \center \includegraphics[scale=0.23]{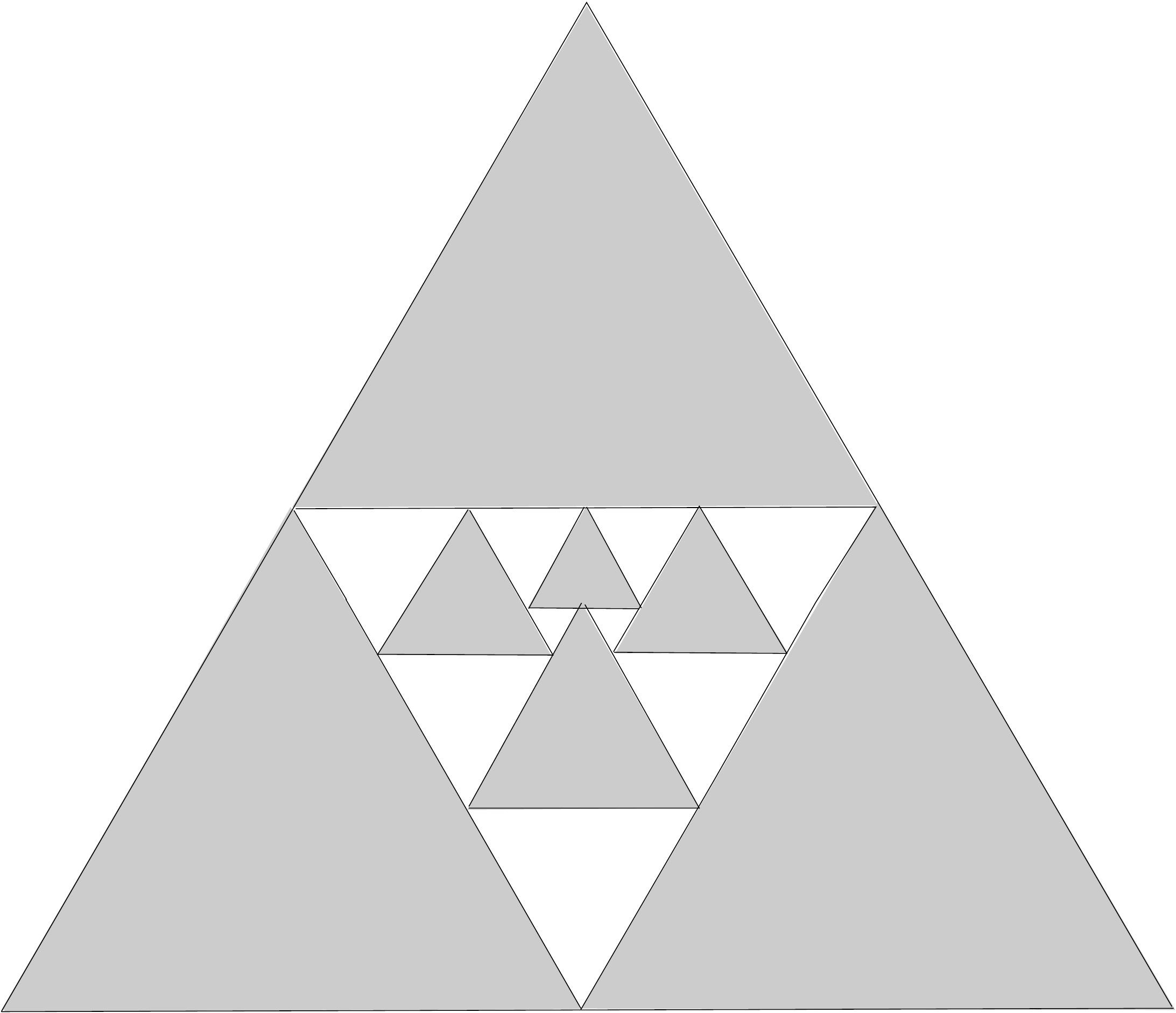}
  \hspace{3em} \includegraphics[scale=0.38]{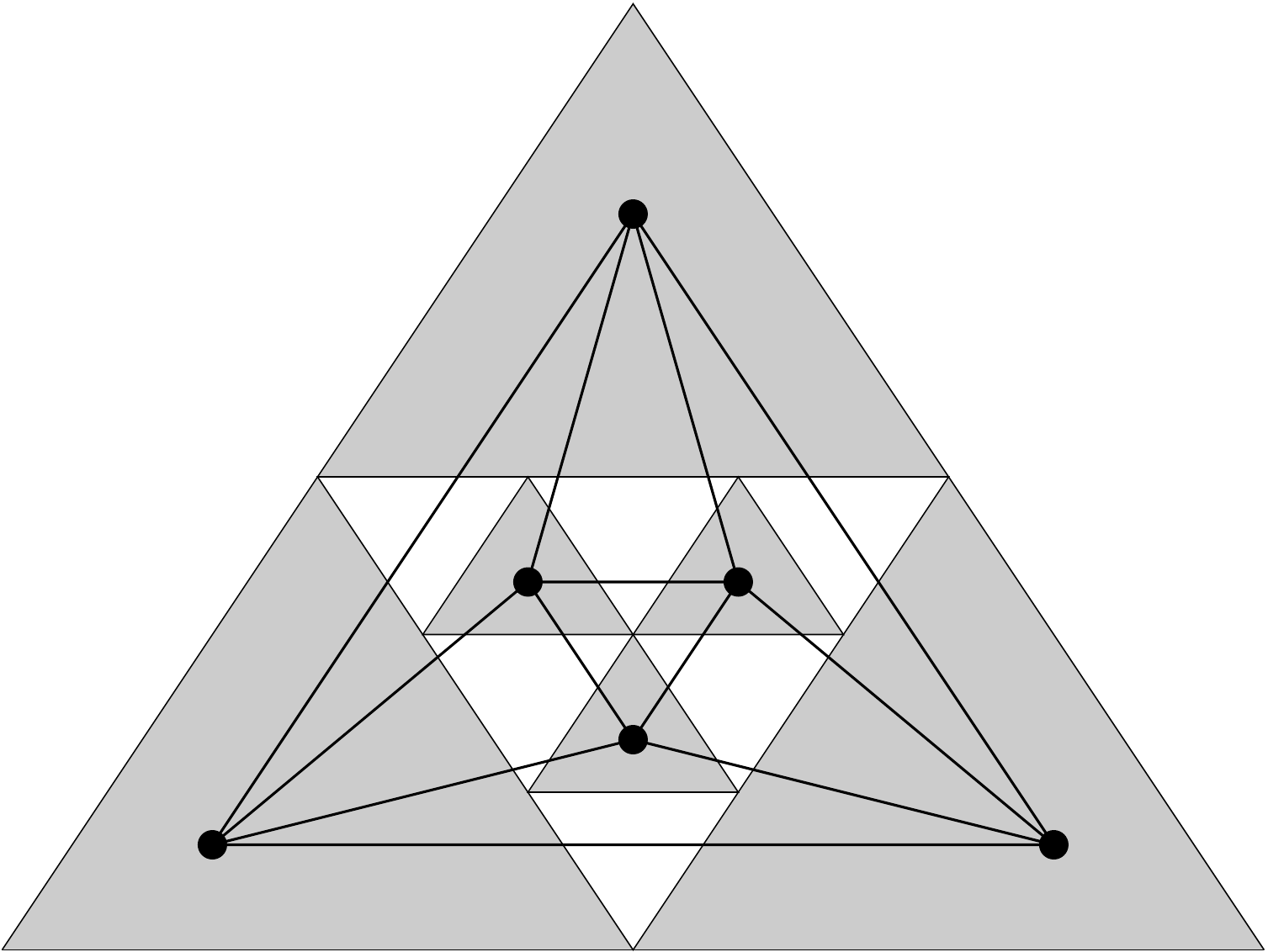}
\caption{Contact representations with homothetic triangles.}
\label{fig:homothetic}
\end{figure}

\begin{theorem}{\cite{GLP11}}
  \label{th:homothetic}
  Every 4-connected planar triangulation admits a contact
  representation with homothetic triangles.
\end{theorem}
Note that one cannot drop the 4-connectedness requirement from
Theorem~\ref{th:homothetic}. Indeed, in every contact representation
of $K_{2,2,2}$ with homothetic triangles, there are three triangles
intersecting in a point (see the right of
Figure~\ref{fig:homothetic}). This implies that the triangulation (not
4-connected) obtained from $K_{2,2,2}$ by adding a degree three vertex
in every face does not admit a contact representation with homothetic
triangles.  Some questions related to this theorem remain open. For
example, it is believed that if a triangulation $T$ admits a contact
representations with homothetic triangles, it is unique up to some
choice for the triangles in the outer-boundary. However this statement
is still not proved.  Another line of research lies in giving another
proof to Theorem~\ref{th:homothetic} (a combinatorial one), or in
providing a polynomial algorithm constructing such a
representation~\cite{Fel-manuscript,Swg17}.

Theorem~\ref{th:homothetic} has a nice consequence. It allowed
Felsner and Francis~\cite{Fel11} to prove that every planar graph has a
contact representation with cubes in $\RR^3$.  In the present paper we
remain in the plane. Theorem~\ref{th:homothetic} is the building block
for proving our main result.  An intersection representation is said
\emph{simple} if every point belongs to at most two shapes.

\begin{theorem}\label{thm:inter}
A graph is planar if and only if it has a simple intersection
representation with homothetic triangles.
\end{theorem}
This answers a conjecture of Lehmann that planar graphs are
max-tolerance graphs (as max-tolerance graphs have shown to be exactly
the intersection graphs of homothetic triangles~\cite{KKLS06}).
M\"uller et al. \cite{MLL11} proved that for some planar graphs, if
the triangle corners have integer coordinates, then their intersection
representation with homothetic triangles needs coordinates of order
$2^{\Omega(n)}$, where $n$ is the number of vertices. The following
section is devoted to the proof of Theorem~\ref{thm:inter}.

\section{Intersection representations with homothetic triangles}\label{sec:homo-inter}

It is well known that simple contact representations produce planar
graphs. The following lemma is slightly stronger.
\begin{lemma}
\label{lem:simple_is_plane}
Consider a graph $G=(V,E)$ given with a simple intersection
representation $\mathcal C=\{c(v) : v\in V\}$. If the shapes $c(v)$
are homeomorphic to disks, and if for any couple $(u,v) \in V^2$ the
set $c(u)\setminus c(v)$ is non-empty and connected, then $G$ is
planar.
\end{lemma}
\begin{proof-}
  Observe that since $\mathcal C$ is simple, the sets $c^\circ(u) =
  c(u)\setminus \left(\cup_{v\in V\setminus \{u\}}c(v)\right)$ are
  disjoint non-empty connected regions.  Let us draw $G$ by first
  choosing a point $p_u$ inside $c^\circ(u)$, for representing each
  vertex $u$. Then for each neighbor $v$ of $u$, draw a curve inside
  $c^\circ(u)$ from $p_u$ to the border of $c(u)\cap c(v)$ (in the
  border of $c^\circ(u)$) to represent the half-edge of $uv$ incident
  to $u$. As the regions $c^\circ(u)$ are disjoint and connected, this
  can be done without crossings. Finally, for each edge $uv$ it is
  easy to link its two half edges by drawing a curve inside $c(u)\cap
  c(v)$. As the obtained drawing has no crossings, the lemma follows.
\end{proof-}

Note that for any two homothetic triangles $\Delta$ and $\Delta'$, the
set $\Delta\setminus \Delta'$ is
connected. Lemma~\ref{lem:simple_is_plane} thus implies the
sufficiency of Theorem~\ref{thm:inter}.  For proving
Theorem~\ref{thm:inter} it thus suffices to construct an intersection
representation with homothetic triangles for any planar graph $G$. In
fact we restrict ourselves to (planar) triangulations because any such
$G$ is an induced subgraph of a triangulation $T$ (an intersection
representation of $T$ thus contains a representation of $G$).  The
following Proposition~\ref{prop:inter-homo} thus implies
Theorem~\ref{thm:inter}.

\begin{figure}[h]
\center
\begin{tabular}{ccc}
  \includegraphics[scale=0.4]{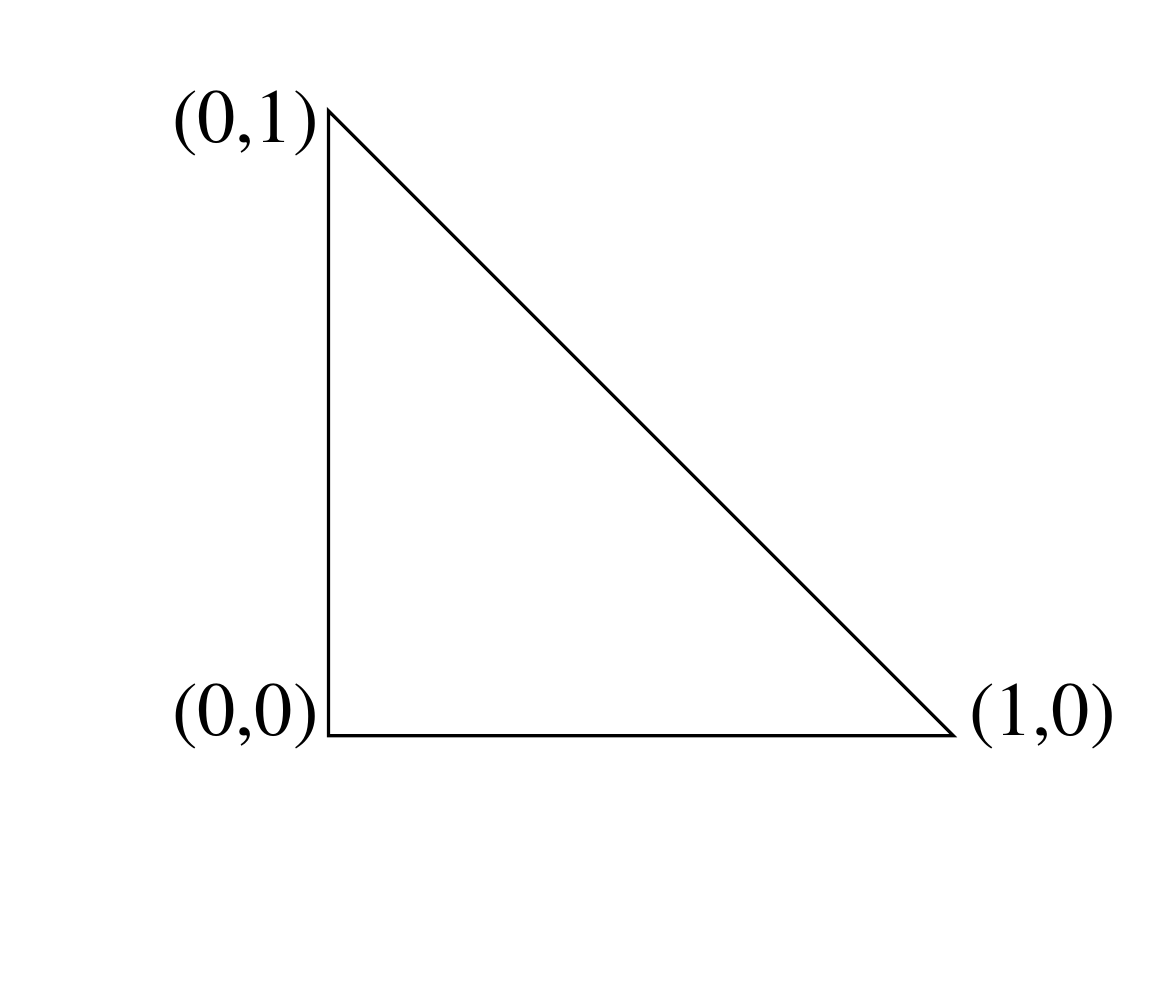}
  & \hspace{4em} &
  \includegraphics[scale=0.4]{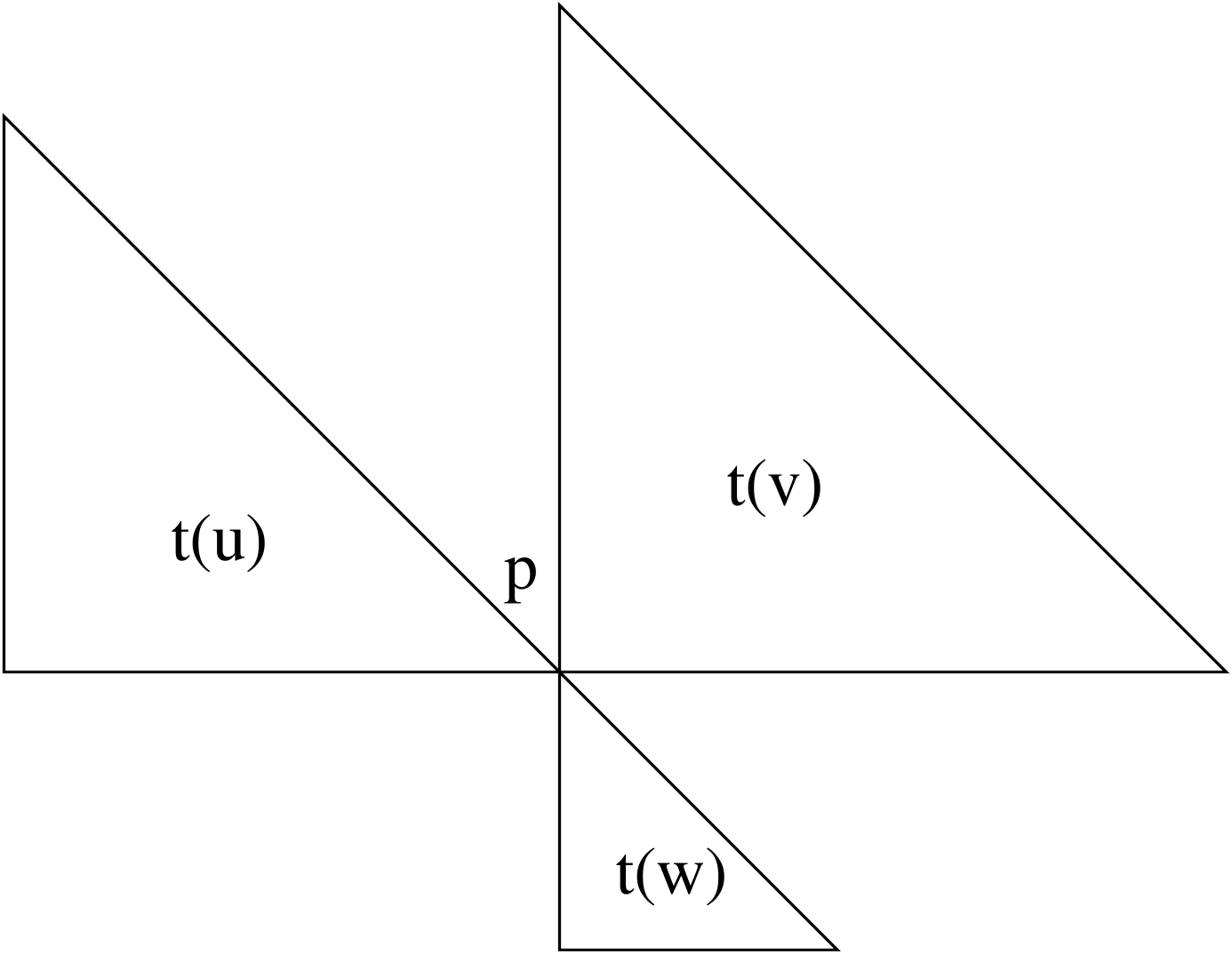}\\
  (a)& &(b) 
\end{tabular}
\hspace{4em}
\caption{(a) The triangle $\Delta$ (b) The triangles $t(u)$, $t(v)$
  and $t(w)$.}
\label{fig:Delta}%{fig:step-0}
\end{figure}

From now on we consider a particular triangle. Given a Cartesian
coordinate system, let $\Delta$ be the triangle with corners at
coordinates $(0,0)$, $(0,1)$ and $(1,0)$ (see
Figure~\ref{fig:Delta}.(a)). Thus the homothets of $\Delta$ have
corners of the form $(x,y)$, $(x,y+h)$ and $(x+h,y)$ with $h>0$, and
we call $(x,y)$ their \emph{right corner} and $h$ their \emph{height}.

\begin{proposition}
\label{prop:inter-homo}
For any triangulation $T$ with outer vertices $a$, $b$ and $c$, for
any three triangles $t(a)$, $t(b)$, and $t(c)$ homothetic to $\Delta$,
that pairewise intersect but do not intersect (i.e. $t(a) \cap t(b)
\cap t(c) = \emptyset$), and for any $\epsilon > 0$, there exists an
intersection representation $\mathcal T=\{t(v) : v\in V(T)\}$ of $T$
with homothets of $\Delta$ such that:
\begin{itemize}
\item[(a)] No three triangles intersect.
\item[(b)] The representation is bounded by $t(a)$, $t(b)$, and $t(c)$
  and the inner triangles intersecting those outer triangles intersect
  them on a point or on a triangle of height less than $\epsilon$.
\end{itemize}
\end{proposition}
\begin{proof-}
Let us first prove the proposition for 4-connected triangulations.
Theorem~\ref{th:homothetic} tells us that 4-connected triangulations
have such a representation if we relax condition (a) by allowing 3
triangles $t(u)$, $t(v)$ and $t(w)$ to
intersect if they pairewise intersect in the same single point $p$
(i.e. $t(u) \cap t(v) = t(u) \cap t(w)
= t(v) \cap t(w) = p$). We call (a') this relaxation
of condition (a), and we call ``bad points'', the points at the
intersection of 3 triangles. Let us now reduce their number (to zero)
as follows (and thus fulfill condition (a)).

Note that the corners of the outer triangles do not intersect
inner triangles. This property will be preserved along the
construction below.

Let $p=(x_p,y_p)$ be the highest (i.e. maximizing $y_p$) bad point. If
there are several bad points at the same height, take among those the
leftmost one (i.e. minimizing $x_p$).  Then let $t(u)$,
$t(v)$ and $t(w)$ be the three triangles
pairewise intersecting at $p$.  Let us denote the coordinates of their
right corners by $(x_u,y_u)$, $(x_v,y_v)$ and $(x_w,y_w)$, and their
height by $h_u$, $h_v$ and $h_w$.  Without loss of generality we let
$p=(x_u+h_u,y_u) = (x_v,y_v) = (x_w,y_w+h_w)$ (see
Figure~\ref{fig:Delta}.(b)). By definition of $p$ it is clear that $p$
is the only bad point around $t(u)$. Note also that none of
$t(u)$, $t(v)$ and $t(w)$ is an
outer triangle.

\begin{figure}[h]
\center
\begin{tabular}{ccc}
  \includegraphics[scale=0.4]{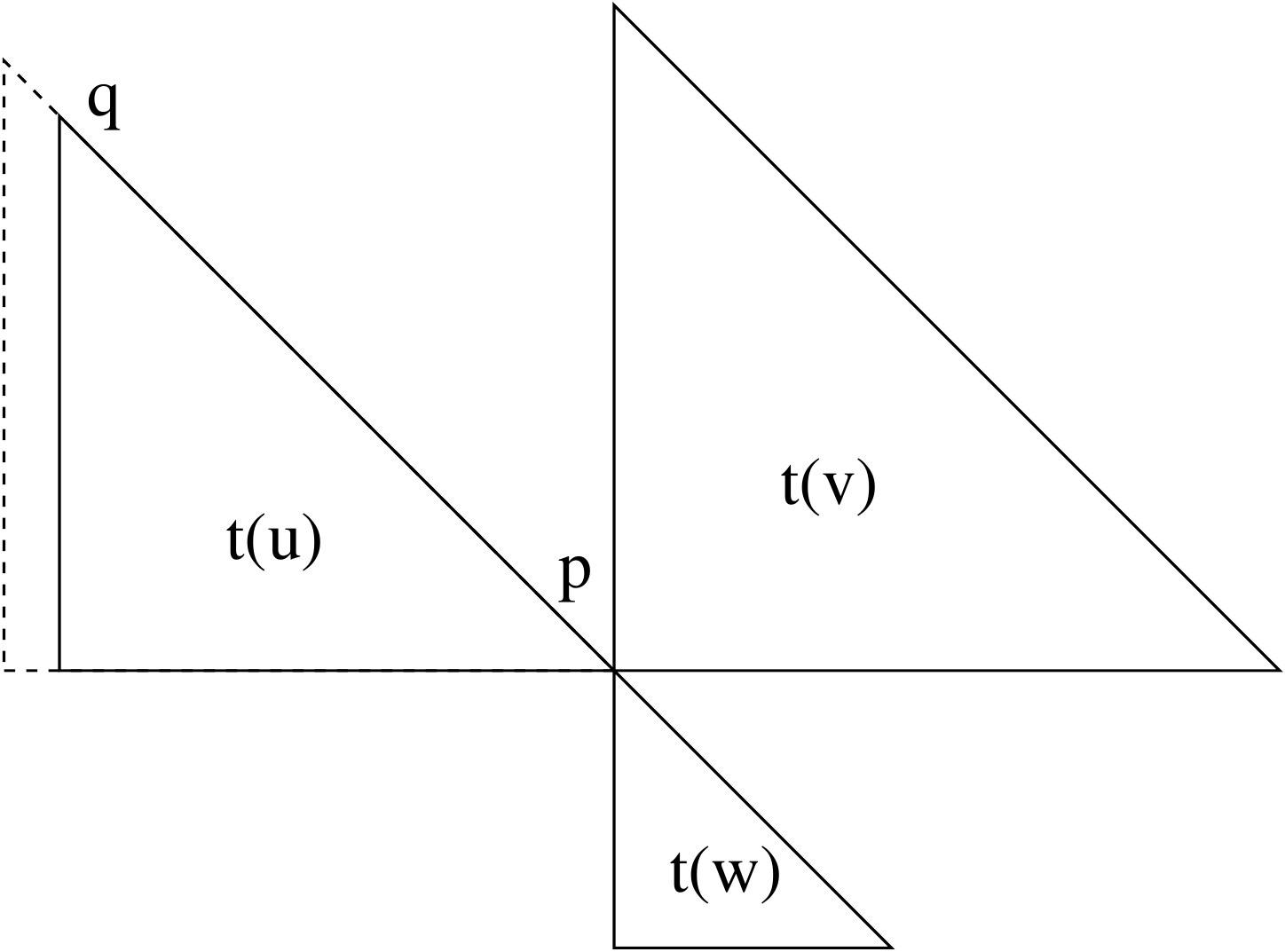}
& \hspace{4em} &
\includegraphics[scale=0.4]{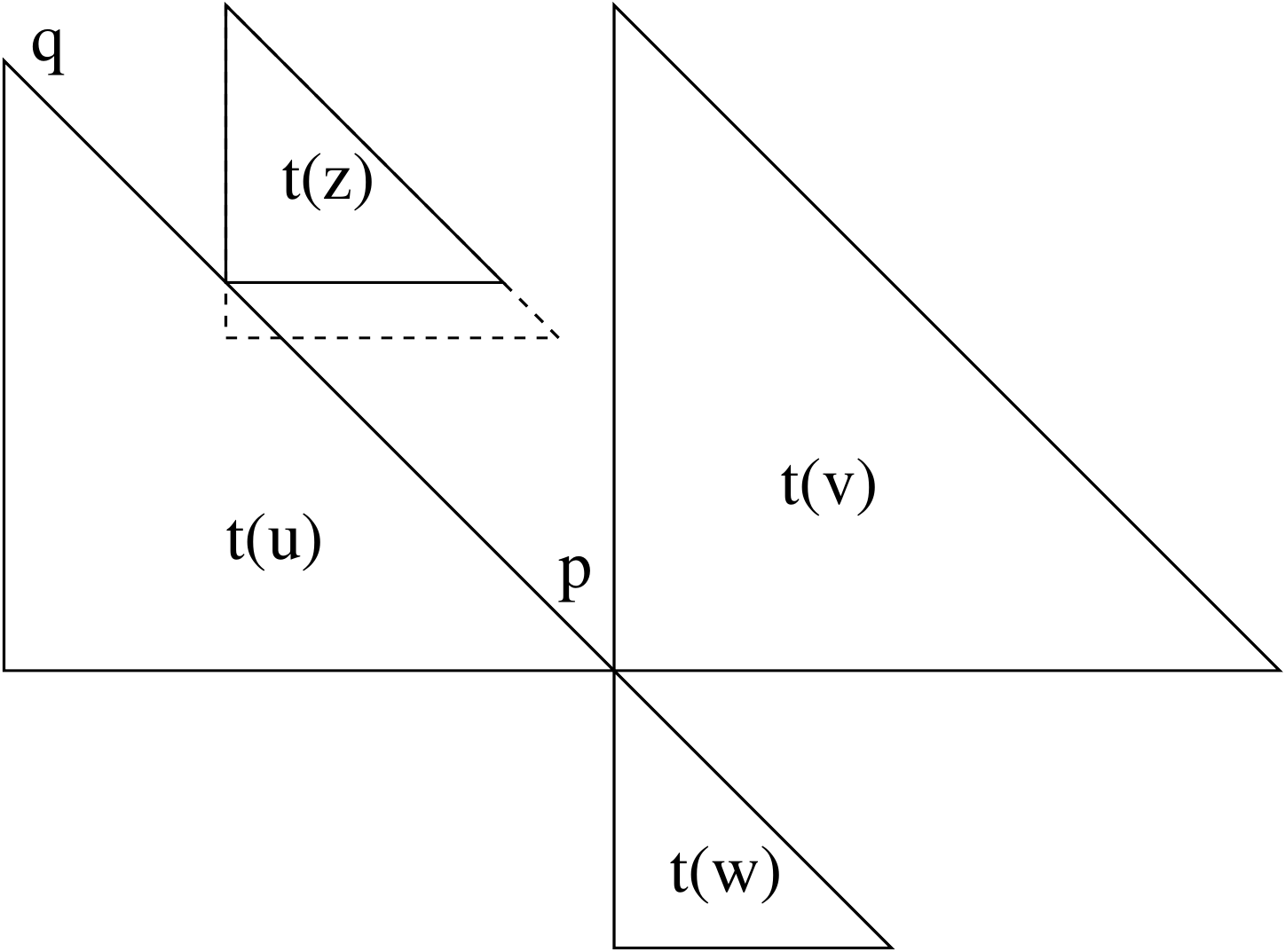}\\
(a)& &(b) 
\end{tabular}
\hspace{4em}
\caption{(a) Step 1 (b) Step 2}
\label{fig:step-12}
\end{figure}

\paragraph{Step 1:}
By definition of $p$ and $t(u)$, the corner $q=(x_u,y_u+h_u)$ of
$t(u)$ is not a bad point.  Now inflate $t(u)$ in order to have its
right angle in $(x_u-\epsilon_1,y_u)$ and height $h_u+\epsilon_1$, for a
sufficiently small $\epsilon_1>0$ (see
Figure~\ref{fig:step-12}.(a)). Here $\epsilon_1$ is sufficiently small
to avoid new pairs of intersecting triangles, new triples of
intersecting triangles, or an intersection between $t(u)$ and an outer
triangle on a too big triangle (with height $\ge \epsilon$).  Since
the new $t(u)$ contains the old one, the triangles originally
intersected by $t(u)$ are still intersected. Hence, $t(u)$ intersects
the same set of triangles, and the new representation is still a
representation of $T$.  Since there was no bad point distinct from $p$
around $t(u)$, it is clear by the choice of $\epsilon_1>0$ that the new
representation still fulfills (a') and (b). After this step we have
the following.
\begin{claim}
The top corner of $t(u)$ is not a contact point.
\end{claim}

\paragraph{Step 2:}
For every triangle $t(z)$ that intersects $t(u)$
on a single point of the open segment $]p,q[$ do the following. Denote
$(x_z,y_z)$ the right corner of $t(z)$, and $h_z$ its
height.  Note that $t(z)$ is an inner triangle of the representation
and that by definition of $p$ there is no bad point involving
$t(z)$. Now inflate $t(z)$ in order to have
its right corner at $(x_z,y_z-\epsilon_2)$, and height $h_z+\epsilon_2$,
for a sufficiently small $\epsilon_2>0$ (see
Figure~\ref{fig:step-12}.(b)). Here $\epsilon_2$ is again sufficiently small
to avoid  new pairs or  new triples of intersecting triangles, and to
preserve (b). Since $t(z)$ was not involved in a bad point,
the new representation still fulfills (a').  Since the new $t(z)$
contains the old one, the triangles originally intersected by
$t(z)$ are still intersected. Hence, $t(z)$
intersects the same set of triangles, and the new representation is still a
representation of $T$.  After doing this to every $t(z)$ we have the
following.
\begin{claim}
There is no contact point on $]p,q]$.
\end{claim}

\begin{figure}[h]
\center
\begin{tabular}{ccc}
\includegraphics[scale=0.4]{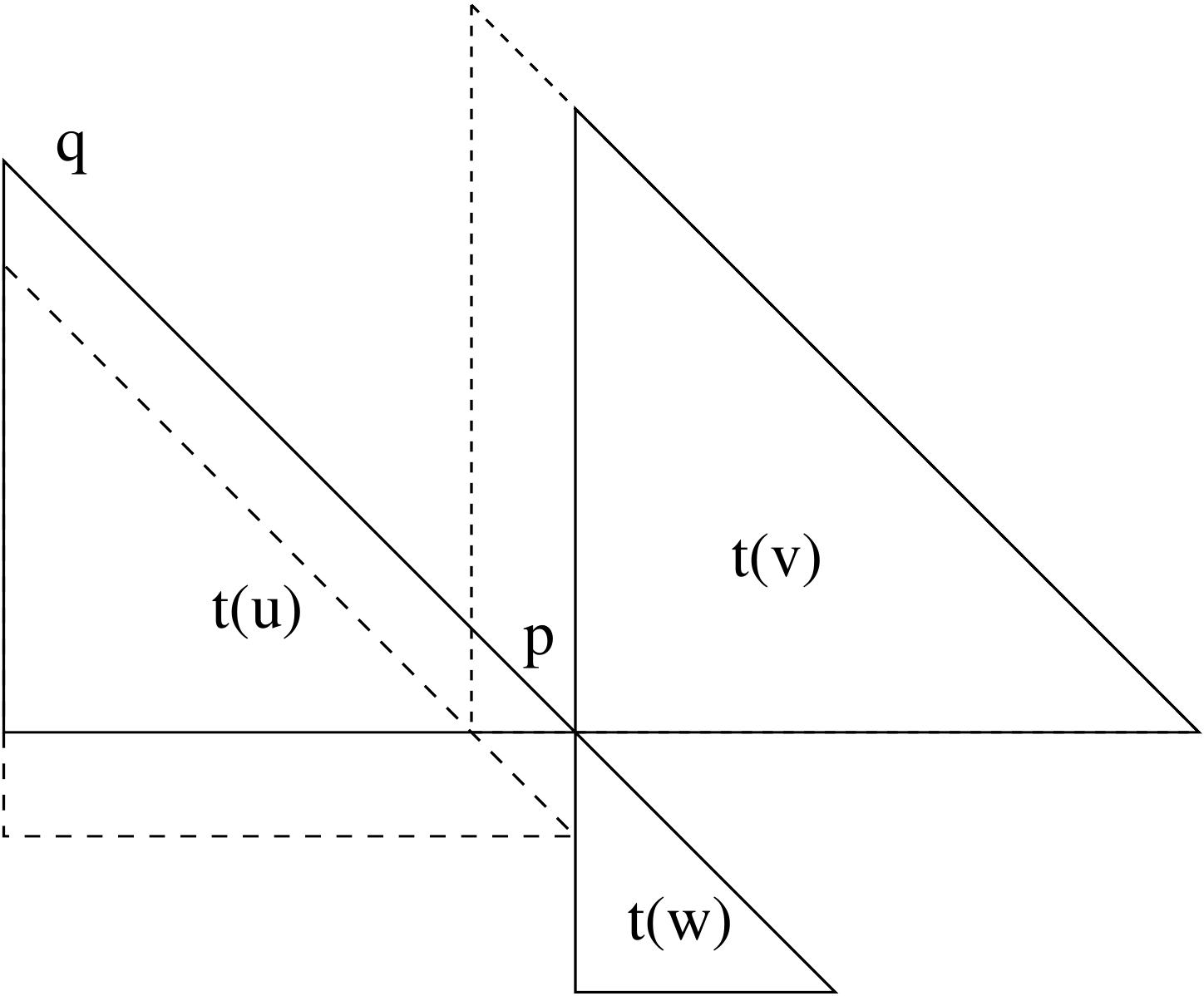}
& \hspace{4em} &
\includegraphics[scale=0.4]{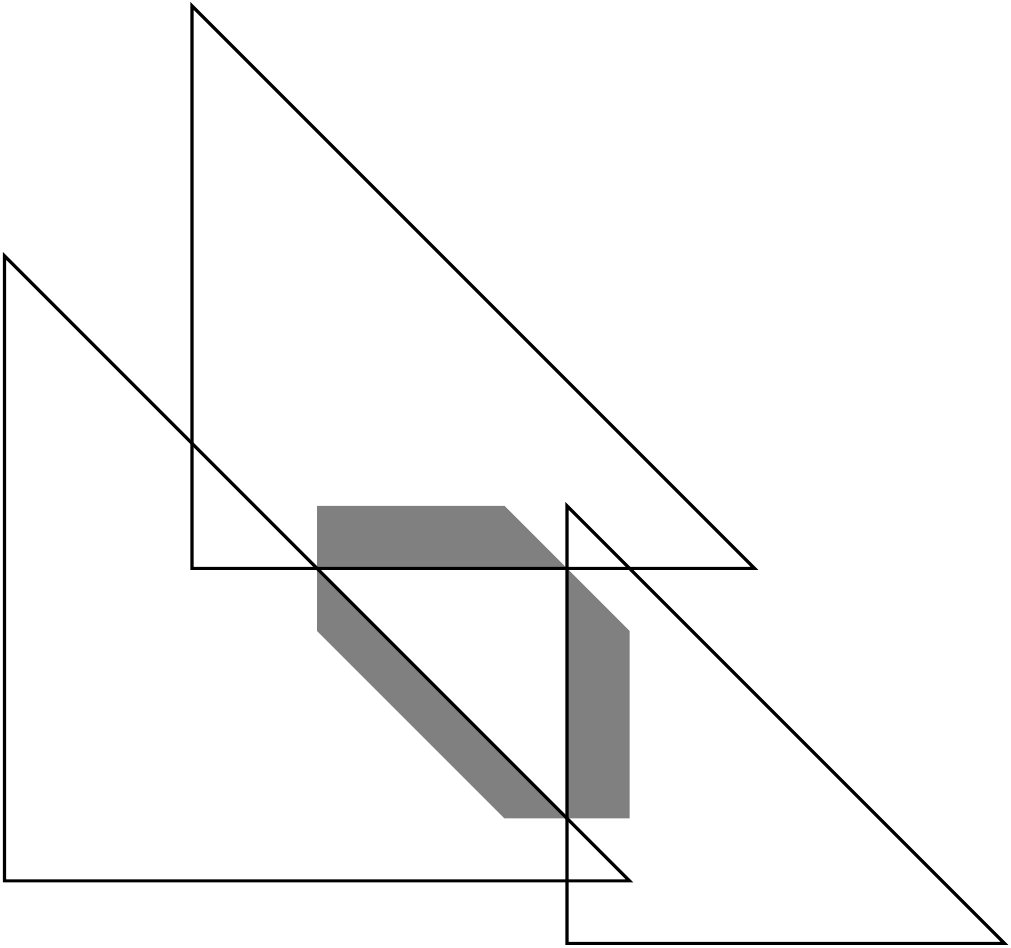}\\
(a)& &(b)
\end{tabular}
\hspace{4em}
\caption{(a) Step 3 (b) Condition (c)}
\label{fig:step-34}
\end{figure}

\paragraph{Step 3:}
Now translate $t(u)$ downwards in order to have its right corner in
$(x_u,y_u-\epsilon_3)$, and inflate $t(v)$ in order to have its right
angle in $(x_v-\epsilon_3,y_v)$, and height $h_v+\epsilon_3$, for a
sufficiently small $\epsilon_3>0$ (see
Figure~\ref{fig:step-34}.(a)). Here $\epsilon_3$ is again sufficiently
small to avoid new pairs or triples of intersecting triangles, and to
preserve (b) but it is also sufficiently small to preserve the
existing pairs of intersecting triangles. This last requirement can be
fulfilled because the only intersections that $t(u)$ could loose would
be contact points on $]p,q]$, which do not exist.

After these three steps, it is clear that the new representation has one bad
point less and induces the same graph. This proves the proposition for
4-connected triangulations.  The conditions (a) and (b) imply the
following property.

\begin{itemize}
\item[(c)] For every inner face $xyz$ of $T$, there exists a triangle
  $t({xyz})$, negatively homothetic to $\Delta$, which interior is
  disjoint to any triangle $t(v)$ but which 3 sides are respectively
  contained in the sides of $t({x})$, $t({y})$ and
  $t(z)$. Furthermore, there exists an $\epsilon' >0$ such that any
  triangle $t$ homothetic to $\Delta$ of height $\epsilon'$ with a
  side in $t({x}) \cap t({xyz})$ does not intersect any triangle
  $t(v)$ with $v\neq x$, and similarly for $y$ and $z$ (see
  Figure~\ref{fig:step-34}.(b) where the grey regions represent the
  union of all these triangles).
\end{itemize}

We are now ready to prove the proposition for any triangulation
$T$. We prove this by induction on the number of separating
triangles. We just proved the initial case of that induction, when $T$
has no separating triangle (i.e. when $T$ is 4-connected). For the
inductive step we consider a separating triangle $(u,v,w)$ and we call
$T_{in}$ (resp. $T_{out}$) the triangulation induced by the edges on
or inside (resp. on or outside) the cycle $(u,v,w)$. By induction
hypothesis $T_{out}$ has a representation fulfilling (a), (b), and
(c). Here we choose arbitrarily the outer triangles and
$\epsilon$. Since $uvw$ is an inner face of $T_{out}$ there exists a
triangle $t({uvw})$ and an $\epsilon' >0$ (with respect to the
inner face $uvw$) as described in (c). Then it suffices to apply the
induction hypothesis for $T_{in}$ (which outer vertices are $u$, $v$
and $w$), with the already existing triangles $t(u)$, $t(v)$, and
$t(w)$ , and for $\epsilon'' = \min(\epsilon,\epsilon')$.  Then one
can easily check that the obtained representation fulfills (a), (b),
and (c). This completes the proof of the proposition.
\end{proof-}

\section{Conclusion}

Given a graph $G$ its \emph{incidence poset} is defined on $V(G)\cup
E(G)$ and it is such that $x$ is greater than $y$ if and only if $x$
is an edge with an end at $y$.  A \emph{triangle poset} is a poset
which elements correspond to homothetic triangles, and such that $x$
is greater than $y$ if and only if $x$ is contained inside $y$. It has
been shown that a graph is planar if and only if its incidence poset
is a triangle poset~\cite{Sch89}\footnote{Triangle posets are exactly
  dimension three posets.}. Theorem~\ref{thm:inter} improves on this
result. Indeed, in the obtained representation the triangles $t(u)$
corresponding to vertices intersect only if those vertices are
adjacent, and the triangles corresponding to edges $uv$, $t(u)\cap
t(v)$, are disjoint.

In $\RR^3$, one can define \emph{tetrahedral posets} as those which
elements correspond to homothetic tetrahedrons in $\RR^3$, and such
that $x$ is greater than $y$ if and only if $x$ is contained inside
$y$. Unfortunately, graphs whose incidence poset is tetrahedral do not
always admit an intersection representation in $\RR^3$ with homothetic
tetrahedrons.  This is the case for the complete bipartite graph
$K_{n,n}$, for a sufficiently big $n$. It is easy to show that its
incidence poset is tetrahedral.  In an intersection representation
with homothetic tetrahedrons, let us prove that the smallest
tetrahedron $t$ has a limited number of neighbors that induce a stable
set.  Let $t'$ be the tetrahedron centered at $t$ and with three times
its size. Note that every other tetrahedron intersecting $t$,
intersects $t'$ on a tetrahedron at least as big as $t$. The limited
space in $t'$ implies that one cannot avoid intersections among the
neighbors of $t$, if they are too many.  The interested reader will
see in~\cite{GI18} that these graphs defined by tetrahedral incidence
posets also escape a characterization as TD-Delaunay graphs.

\bibliographystyle{plain}

\end{document}